\documentclass[aip,reprint]{revtex4-1}

\usepackage[T1]{fontenc} 
\usepackage{natbib}
\usepackage{amsmath}
\usepackage{graphicx}

\newcommand{\esref}[1]{Eqs.~(\ref{#1})}
\newcommand{\fref}[1]{Fig.~\ref{#1}}

\begin{document}


\title{Semiconductor Ring Lasers Coupled by a Single Waveguide}


\author{W. Coomans}
\email[Electronic address: ]{wcoomans@vub.ac.be}
\author{L. Gelens}
\author{G. Van der Sande}
\affiliation{Applied Physics Research Group, Vrije Universiteit Brussel, Pleinlaan 2, B-1050 Brussel, Belgium}
\author{G. Mezosi}
\author{M. Sorel}
\affiliation{Department of Electronics and Electrical Engineering, University of Glasgow, Oakfield Avenue, Glasgow G12 8LT, Scotland, UK}
\author{J. Danckaert}
\author{G. Verschaffelt}
\affiliation{Applied Physics Research Group, Vrije Universiteit Brussel, Pleinlaan 2, B-1050 Brussel, Belgium}



\date{\today}

\begin{abstract}
We experimentally and theoretically study the characteristics of semiconductor ring lasers bidirectionally coupled by a single bus waveguide. This configuration has, e.g., been suggested for use as an optical memory and as an optical neural network motif. The main results are that the coupling can destabilize the state in which both rings lase in the same direction, and it brings to life a state with equal powers at both outputs. These are both undesirable for optical memory operation. Although the coupling between the rings is bidirectional, the destabilization occurs due to behavior similar to an optically injected laser system.
\end{abstract}


\maketitle


Semiconductor ring lasers (SRLs) are characterized by a cavity with a circular geometry. As a result, SRLs can generate light in two counterpropagating directions referred to as the clockwise (CW) and the counterclockwise (CCW) mode. This bistable character enables them to be used in systems for all-optical switching and as all-optical memories, both in solitary \cite{Sorel_APL_2002,Liu_NatPhot_2010} and coupled \cite{Hill_Nature_2004,Ishii_JQE_2006,DeConinck_LEOS_2009,Lin_OL_2011} configurations.
The convenient device properties of SRLs allow them to be highly integrable and scalable,\cite{Krauss_EL_1990} making them suitable candidates for key components in photonic integrated circuits.
In a broader context, the topic of coupled lasers received much attention in recent years.\cite{Erzgraber_PRE_2010,Brunstein_APL_2011,Bonatto_PRE_2012}

In this contribution, we investigate two SRLs that are bidirectionally coupled through a single bus waveguide. Hence, only one of the counterpropagating modes in each SRL receives direct input from the other SRL, which introduces an asymmetry in the global configuration. This coupling scheme has already been suggested for use as an optical memory\cite{Hill_Nature_2004,DeConinck_LEOS_2009} and as an optical neural network motif using asymmetric (excitable) SRLs.\cite{Coomans_PRE_2011}
In both types of applications, it is important to investigate the influence of the coupling on the behavior of the device.

The experiments have been performed on InP-based multi-quantum-well SRLs with a racetrack geometry. The device is mounted on a copper mount and is thermally controlled by a Peltier element that is stabilized with an accuracy of 0.01$^\circ$C.
We used a fast photodiode with a 12.5 GHz bandwidth coupled to an Anritsu MS2667C 30 GHz RF spectrum analyzer. Optical spectra have been measured with a grating based OSA (Ando AQ6317B), using a resolution of 0.02 nm.
\begin{figure}[tb!]
\begin{center}
\includegraphics{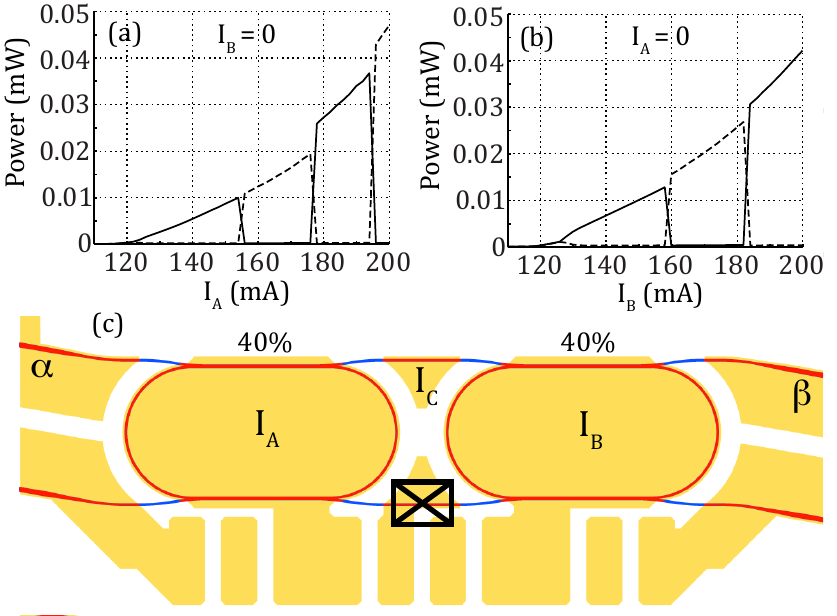}
\caption{(Color online) (a),(b) PI curve of respectively SRL A and SRL B. The CW (CCW) mode is indicated by a solid (dashed) line. (c) Chip layout with its contacting. SRL A (B) is biased with a current $I_\mathrm{A}$ ($I_\mathrm{B}$). The coupling waveguide can be biased with a current $I_\mathrm{C}$. The bottom waveguide doesn't guide any light from one laser to the other.}
\label{fig:chiplayout}
\end{center}
\end{figure}

The chip layout is shown in \fref{fig:chiplayout}(c). There are two separate SRLs A and B (bias currents $I_\mathrm{A}$ and $I_\mathrm{B}$), whose output power is collected through evanescent coupling to an output waveguide. This output waveguide also serves as the coupling waveguide and can be biased independently ($I_\mathrm{C}$) to control the coupling strength. Both SRLs and all waveguides are fabricated from the same active material.
Note that the bottom coupling waveguide has been interrupted and does not guide any light from one laser to the other. Hence, the SRLs are only coupled through the top waveguide in \fref{fig:chiplayout}(c).
PI curves of SRL A and B are shown in \fref{fig:chiplayout}(a) and (b).
The threshold current of both SRL A and B is measured to be 120~mA, and the longitudinal mode spacing is 0.41~nm (51~GHz).
Very close to threshold, the SRLs operate bidirectionally with equal power in the CW and CCW mode. For currents larger than 125 mA they operate unidirectionally, in which one of the counterpropagating modes dominates over the other.
The relaxation oscillation (RO) frequency of the SRLs is about 4 GHz at a bias current of 175 mA. The free running wavelength of SRL B is a couple of nm ($\approx$ 2 to 4~nm) larger than the free running wavelength of SRL A, depending on the longitudinal mode that is selected to lase.

Solitary SRLs operating in the unidirectional regime can lase in either the CW or the CCW direction.\cite{Gelens_PRL_2009} This gives rise to 4 different combinations for the coupled system: both SRLs can either lase in the same or in the opposite direction. We will refer to the state in which the SRLs lase in the same direction as the asymmetric state, and to the state in which the SRLs lase in opposite direction as the symmetric state.

\begin{figure}[tb!]
\begin{center}
\includegraphics{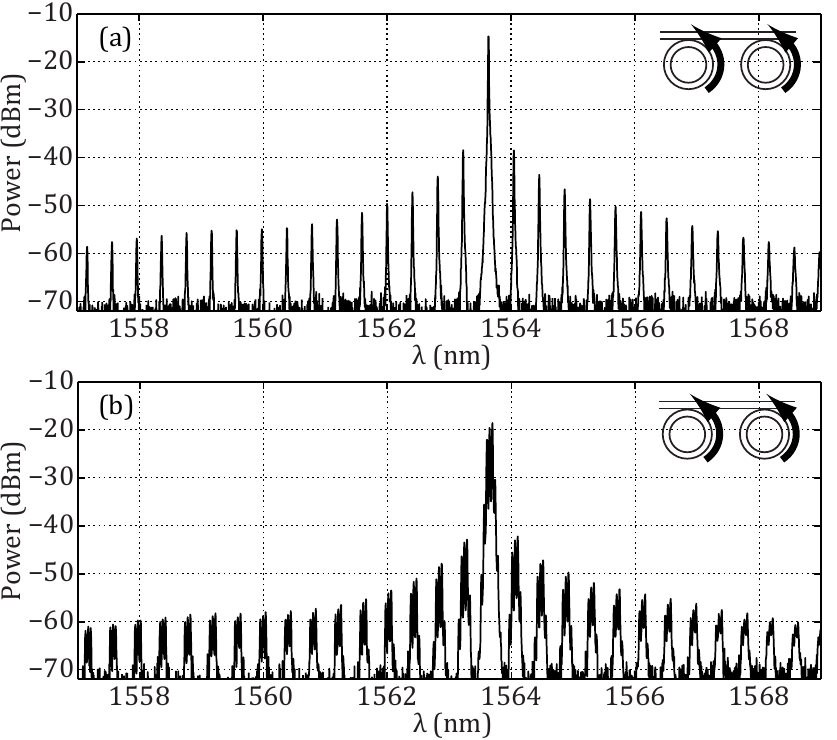}
\caption{Optical spectra of the asymmetric state. (a) $I_\mathrm{C}=0~\mathrm{mA}$, locked state. (b) $I_\mathrm{C}=10~\mathrm{mA}$, destabilized locked state. $I_\mathrm{A}=176~\mathrm{mA}$, $I_\mathrm{B}=169~\mathrm{mA}$. Measurements at $\alpha$ port.}
\label{fig:OSA}
\end{center}
\end{figure}
When the SRLs are in the asymmetric state, locking can be achieved by aligning the longitudinal mode combs through a positive offset $\Delta I=I_\mathrm{A}-I_\mathrm{B}$ in bias current. Both SRLs then lase in the same longitudinal mode (see \fref{fig:OSA}(a)). There is a tolerance of a few mA on the offset current to obtain locked operation.
The lasing wavelength of the locked state is determined by a master-slave relationship. If both SRLs are lasing towards the $\alpha$ ($\beta$) port, SRL A (B) acts as the slave, and SRL B (A) controls the lasing wavelength.\cite{Hill_Nature_2004}
When increasing the coupling strength (increasing $I_\mathrm{C}$) the asymmetric locked state is destabilized and we observe a modulation of the optical spectrum (see \fref{fig:OSA}(b)). The fact that all longitudinal modes are modulated in the same way indicates that this is due to gain dynamics. The modulation is sustained for all investigated values of $I_\mathrm{C}$ (i.e.\ up to 50 mA).
This destabilization of the asymmetric locked state for increasing coupling strength happens for all investigated values of the bias currents on the SRLs.

\begin{figure}[tb!]
\begin{center}
\includegraphics{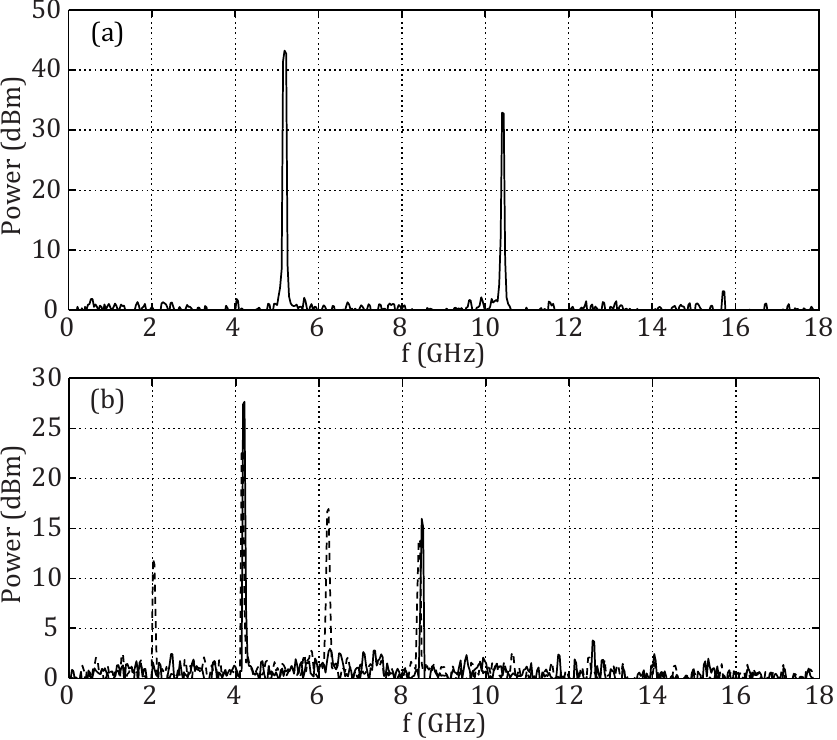}
\caption{RF spectra. (a) $I_\mathrm{A}=176~\mathrm{mA}$, $I_\mathrm{B}=169~\mathrm{mA}$ and $I_\mathrm{C}=10~\mathrm{mA}$. (b) $I_\mathrm{A}=155.5~\mathrm{mA}$, $I_\mathrm{B}=145~\mathrm{mA}$, $I_\mathrm{C}=15~\mathrm{mA}$ (solid) and $I_\mathrm{C}=17~\mathrm{mA}$ (dashed). Measurements at $\alpha$ port.}
\label{fig:RFPD}
\end{center}
\end{figure}

A typical RF spectrum of this regime is shown in \fref{fig:RFPD}(a), indicating a narrow peak at 5.18 GHz and its first harmonic for bias currents $I_\mathrm{A}=176$~mA and $I_\mathrm{B}=169$~mA. Note that this value is of the same order as, but larger than, the RO frequency of the solitary SRLs.
For lower values of the SRL bias currents (e.g., $I_\mathrm{A}=155.5$~mA and $I_\mathrm{B}=145$~mA) we additionally observe a range of values of $I_\mathrm{C}$ over which the oscillation doubles its period (see \fref{fig:RFPD}(b)), after which the period one oscillation is restored.
\begin{figure}[tb!]
\begin{center}
\includegraphics{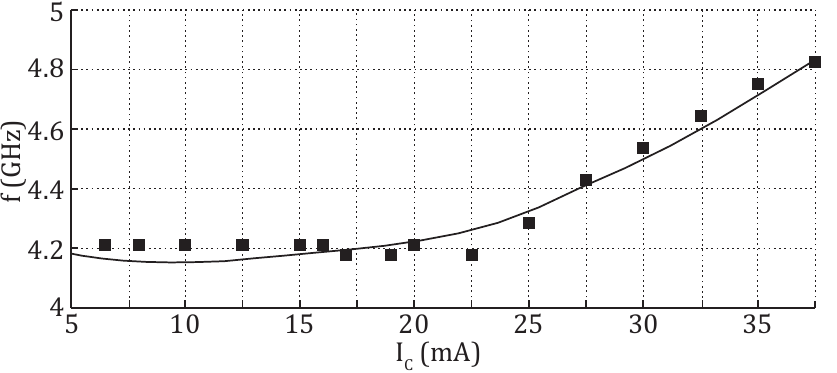}
\caption{Experimental (squares) and simulated (solid line) frequency of the observed oscillations. The coupling amplitude $k_c$ is fitted to the coupling waveguide current $I_\mathrm{C}$ by $I_\mathrm{C}=0.1676(k_c\times\mathrm{ns})^2$ mA. Other simulation parameters as in \fref{fig:simuTT}(b). $I_\mathrm{A}=155.5~\mathrm{mA}$, $I_\mathrm{B}=145~\mathrm{mA}$.}
\label{fig:fvsIc}
\end{center}
\end{figure}

The oscillation frequency increases for increasing bias currents of the SRLs, as can be seen by comparing \fref{fig:RFPD}(a) and (b). For increasing $I_\mathrm{C}$ the oscillation frequency generally increases, except for low bias currents. In that case the frequency first remains constant, and only increases with $I_\mathrm{C}$ from the period doubling on(see squares in \fref{fig:fvsIc}).

\begin{figure}[tb!]
\begin{center}
\includegraphics{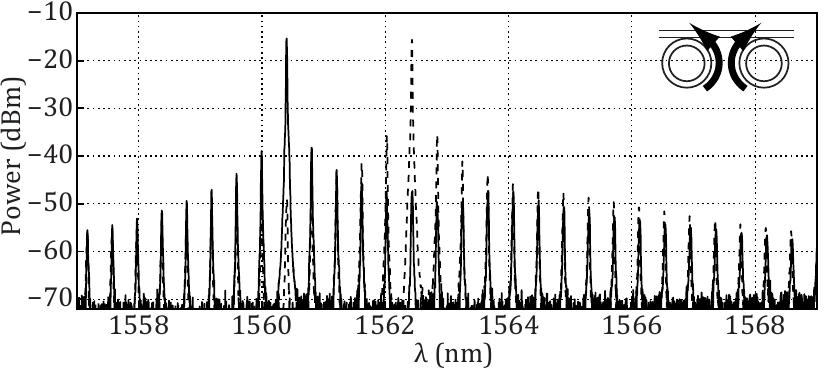}
\caption{Optical spectrum of the symmetric state at the $\alpha$ port (solid) and $\beta$ port (dashed). $I_\mathrm{A}=176~\mathrm{mA}$, $I_\mathrm{B}=169~\mathrm{mA}$ and $I_\mathrm{C}=0~\mathrm{mA}$.}
\label{fig:ab}
\end{center}
\end{figure}

Furthermore, bistability was observed between the asymmetric state and the symmetric state (e.g., see \fref{fig:OSA}(a) and \fref{fig:ab}).
When the SRLs are in the symmetric state, they lase at their own wavelength (see \fref{fig:ab}). Although the frequency combs are aligned, no locking is observed when the SRLs lase in opposite directions, even when greatly increasing the coupling strength through $I_\mathrm{C}$. Note that the global system is operating \emph{bidirectionally}, with equal powers at both output ports, while both SRLs are actually operating \emph{unidirectionally}, with more power in one of the counterpropagating modes than in the other. In this regime, the modes that are coupled to the other SRL are the low power modes. Due to the flatness of the optical spectrum of these low power modes,\cite{Born_JQE_2005} the SRLs are not able to phase-lock. The many longitudinal modes with approximately equal power do not provide a suitable phase reference for the other SRL.

To model the destabilization of the asymmetric locked state, we use a rate equation model\cite{Sorel_OL_2002} assuming single transverse and single longitudinal mode operation and the existence of two counterpropagating modes.
For each SRL X ($\mathrm{X}=\{\mathrm{A},\mathrm{B}\}$), the model consists of two slowly varying complex envelopes of the counterpropagating waves $E_{1\mathrm{X}}$ (CW) and $E_{2\mathrm{X}}$ (CCW) and a third equation for the carrier population inversion $N_\mathrm{X}$.
\begin{subequations}
\label{eq:RateEqCpld}
\small
\begin{align}
\dot E_{1\mathrm{A}} &= \kappa \left( {1 + \mathrm{i}\alpha } \right)\left[g_{1\mathrm{A}} N_\mathrm{A} - 1\right]E_{1\mathrm{A}}
-ke^{\mathrm{i}\phi_k}E_{2\mathrm{A}} \label{eq:RA1a}\\
\dot E_{2\mathrm{A}} &= \kappa \left( {1 + \mathrm{i}\alpha } \right)\left[g_{2\mathrm{A}} N_\mathrm{A} - 1\right]E_{2\mathrm{A}} - ke^{\mathrm{i}\phi_k}E_{1\mathrm{A}} \notag \\&\qquad - k_ce^{\mathrm{i}\phi_c}E_{2\mathrm{B}} \label{eq:RA2a}\\
\dot E_{1\mathrm{B}} &= \kappa \left( {1 + \mathrm{i}\alpha } \right)\left[g_{1\mathrm{B}} N_\mathrm{B} - 1\right]E_{1\mathrm{B}}
-ke^{\mathrm{i}\phi_k}E_{2\mathrm{B}}\notag\\&\qquad - k_ce^{\mathrm{i}\phi_c}E_{1\mathrm{A}} \label{eq:RA1b}\\
\dot E_{2\mathrm{B}} &= \kappa \left( {1 + \mathrm{i}\alpha } \right)\left[g_{2\mathrm{B}} N_\mathrm{B} - 1\right]E_{2\mathrm{B}} - ke^{\mathrm{i}\phi_k}E_{1\mathrm{B}} \label{eq:RA2b} \\
\dot N_\mathrm{A} &= \gamma \left[ \mu - N_\mathrm{A} - g_{1\mathrm{A}}N_\mathrm{A}\left|E_{1\mathrm{A}} \right|^2 - g_{2\mathrm{A}}N_\mathrm{A}\left|E_{2\mathrm{A}} \right|^2\right] \\
\dot N_\mathrm{B} &= \gamma \left[ \mu - N_\mathrm{B} - g_{1\mathrm{B}}N_\mathrm{B}\left|E_{1\mathrm{B}} \right|^2 - g_{2\mathrm{B}}N_\mathrm{B}\left|E_{2\mathrm{B}} \right|^2\right]
\end{align}
\end{subequations}
Here the dot represents differentiation with respect to time, $g_{1\mathrm{X}}=1-s|E_{1\mathrm{X}}|^2-c|E_{2\mathrm{X}}|^2$ and $g_{2\mathrm{X}}=1-s|E_{2\mathrm{X}}|^2-c|E_{1\mathrm{X}}|^2$ are differential gain functions that include phenomenological cross ($c$) and self-saturation ($s$) terms (with $c=2s$), $\mu$ is the renormalized injection current ($\mu=0$ at transparency and $\mu=1$ at lasing threshold), $\kappa$ is the field decay rate, $\gamma$ is the carrier decay rate and $\alpha$ is the linewidth enhancement factor.
Reflections due to imperfections inside the cavity result in a linear coupling between the two counterpropagating fields within each SRL (backscattering), and are modeled by an amplitude $k$ and a phase shift $\phi_k$.
The coupling between the SRLs is modeled phenomenologically by a coupling amplitude $k_c$ and a coupling phase $\phi_c$. 
This phase description of the coupling (neglecting any effects of a delay time) is justified since the travel time between the SRLs is of the same order as the cavity round trip time in the experimental setup. 
The value of the coupling parameters $k_c$ and $\phi_c$ is a priori unknown. We have chosen $\phi_c=0$, but all results discussed below are valid for all values $\phi_c\in[0,\pi]$ (the dynamics of \esref{eq:RateEqCpld} is $\pi$-periodic in $\phi_c$).\cite{Coomans_PRE_2011}
For simplicity, we use identical parameter values for SRL A and B.

\begin{figure}[t!]
\begin{center}
\includegraphics{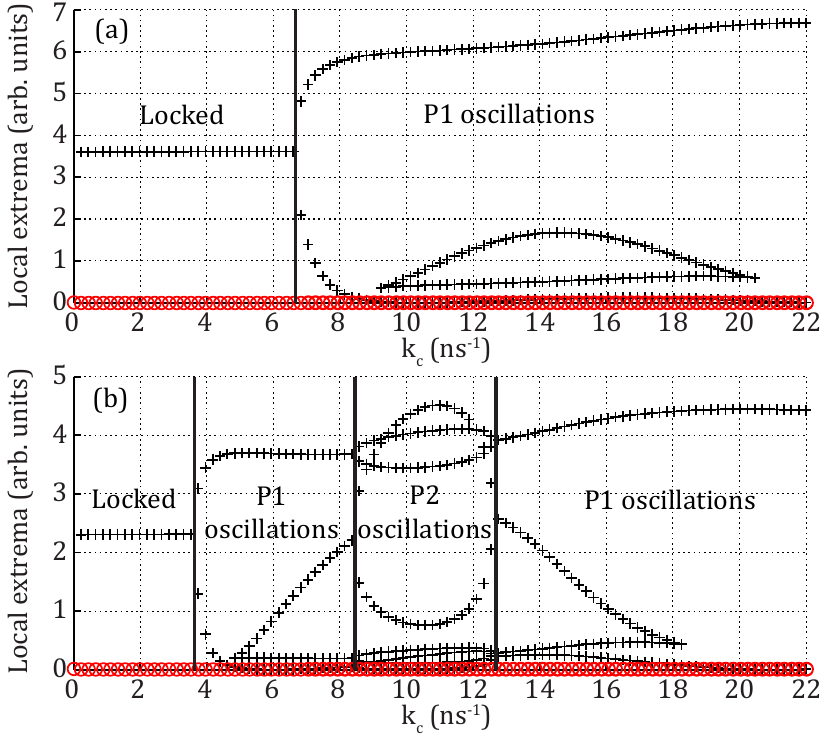}
\caption{(Color online) Numerical simulation of \esref{eq:RateEqCpld} in the asymmetric state. Orbit diagram depicting the local extrema of the power at the $\alpha$ port (black +) and the $\beta$ port (red o) of the chip vs.\ the coupling amplitude $k_c$. (a) $\mu=3.5$. (b) $\mu=2.6$. Other parameters: $\kappa=500$~ns$^{-1}$, $\gamma=0.38$~ns$^{-1}$, $\alpha=3.5$, $s=0.005$, $k=0.44~\mathrm{ns}^{-1}$, $\phi_k=1.5$, $\phi_c=0$.}
\label{fig:simuTT}
\end{center}
\end{figure}

Figures \ref{fig:simuTT}(a) and \ref{fig:simuTT}(b) show orbit diagrams of the total powers emitted at both sides of the chip for two different bias currents, respectively $\mu=3.5$ and $\mu=2.6$. The field at each side of the chip is given by $E_\alpha = E_{2\mathrm{A}}+E_{2\mathrm{B}}\exp(\mathrm{i}\phi_c)$ and $E_\beta = E_{1\mathrm{A}}\exp(\mathrm{i}\phi_c)+E_{1\mathrm{B}}$. For each value of the coupling amplitude $k_c$ they show the local extrema of the total powers $|E_\alpha|^2$ and $|E_\beta|^2$. For low $k_c$ the SRLs reside in the asymmetric locked state, as in the experiment. 
For both bias currents the destabilization of this state for increasing $k_c$ is clearly reproduced, at respectively $k_c\approx6.5$\ ns$^{-1}$ and $k_c\approx3.5$\ ns$^{-1}$. 
This happens through a Hopf bifurcation of the asymmetric locked state. The influence of $\phi_c$ is limited to marginally moving the bifurcation points to a different value of $k_c$. 
Although these are coupled lasers with a bidirectional coupling, the simulations show that during the oscillations in \fref{fig:simuTT} SRL B acts as the master laser with an approximately constant output power and SRL A is the slave laser that is driven into relaxation oscillations, similar to an optical injection system where the coupling is unidirectional.\cite{Coomans_PRA_2010}
For $\mu=2.6$, the additional period doubling of the limit cycle is also reproduced (approximately between $k_c=9$\ ns$^{-1}$ and $k_c=13$\ ns$^{-1}$ in \fref{fig:simuTT}(b)).
Note that the extra local extrema appearing at the bottom in both \fref{fig:simuTT}(a) and (b) are no period doubling, but simply an artifact of the oscillating phase difference between SRL A and B.
\begin{figure}[t!]
\begin{center}
\includegraphics{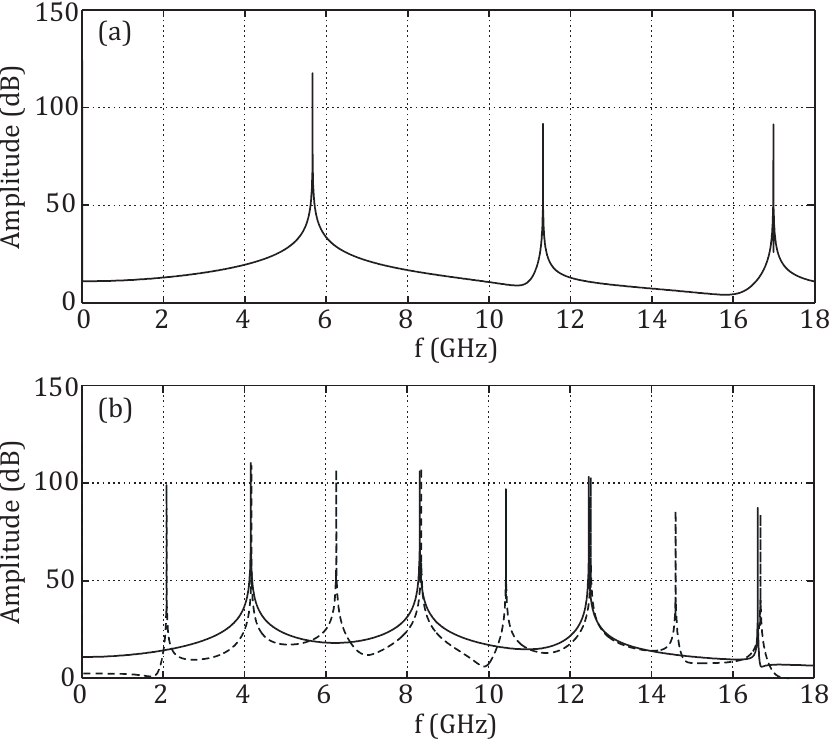}
\caption{Numerical simulation of \esref{eq:RateEqCpld}. RF spectra of the power at the $\alpha$ port. (a) $\mu=3.5$, $k_c=8$\ ns$^{-1}$. (b) $\mu=2.6$, $k_c=8$\ ns$^{-1}$ (solid) and $k_c=9$\ ns$^{-1}$ (dashed). Other parameters as in \fref{fig:simuTT}.}
\label{fig:simuRF}
\end{center}
\end{figure}
We numerically reproduced the experimental RF spectra of \fref{fig:RFPD} in \fref{fig:simuRF}, with good correspondence. 
The value $\mu=3.5$ for \fref{fig:simuRF}(a) was obtained by fitting $\mu$ and $I_\mathrm{A}$ at threshold and at the occurrence of the period doubling. Hence, identifying $\mu=2.6$ with $I_\mathrm{A}=155.5$mA (period doubling) and $\mu=1$ with $I_\mathrm{A}=120$~mA (threshold), yields the approximate relationship $\mu=0.045(I_\mathrm{A}\times\mathrm{mA}^{-1}-120)+1$.

The variation of the oscillation frequency with the coupling strength is numerically reproduced by the solid line in \fref{fig:fvsIc}. 
Note that the formation of carrier gratings in the coupling waveguide and rings is not taken into account in the modeling. Such gratings could contribute to the frequency scaling in \fref{fig:fvsIc}, as shown for self-modulation oscillations in solid-state ring lasers.\cite{Kravtsov_LP_1993} Nevertheless, it was shown that the effect of carrier gratings is much weaker in semiconductor lasers than in solid-state lasers.\cite{Etrich_JQE_1992}
The numerical simulations also predict that the asymmetric locked state regains stability for large coupling strengths (not shown), but it was impossible to investigate this experimentally due to the current limit on the coupling waveguide.
The experimentally observed symmetric state (see \fref{fig:ab}) is also observed in the numerical simulations. But in this single mode model the SRLs lock, such that both sides of the chip lase at identical wavelengths. Modeling including different longitudinal modes, such as a traveling wave model,\cite{Javaloyes_JQE_2009} are needed to reproduce the experimentally observed absence of locking in the symmetric state and also the master-slave relationship which determines the lasing wavelength in the asymmetric state.

In summary, we experimentally and theoretically investigated SRLs that are coupled by a single bus waveguide. When this configuration is used as an optical memory, the coupling between the rings can destabilize the asymmetric locked state (in which both SRLs lase in the same direction and are phase-locked) by exciting relaxation oscillations. This results in a periodic rather than a stable output power level. The coupling also brings to life a symmetric state, i.e. they lase in opposite directions leading to symmetric output power levels. Experiments show that the SRLs cannot phase-lock in this state, even for large coupling strengths. They hence lase at  their own free-running wavelength, which leads to equal powers but different lasing wavelengths at the two output ports.

This work has been funded by the Research Foundation Flanders (FWO). The authors also acknowledge the support of the Belgian Science Policy Office under the IAP project ``photonics@be''. G.M.\ gratefully acknowledges the technical team of the James Watt Nanofabrication Centre for the support in fabricating the devices.

%

\end{document}